# Polycaprolactone/graphite nanoplates composite nanopapers

Kun Li[1], Daniele Battegazzore[2], Orietta Monticelli[1], Alberto Fina[2]

[1] Dipartimento di Chimica e Chimica Industriale, Università di Genova, Via Dodecaneso 31, 16146 Genova, Italy

[2] Dipartimento di Scienza Applicata e Tecnologia, Politecnico di Torino- Alessandria campus, viale Teresa Michel, 5, 15121 Alessandria, Italy

corresponding author: alberto.fina@polito.it

## Abstract

Nanopapers based on graphene and related materials were recently proposed for application in heat spreader applications. To overcome typical limitations in brittleness of such materials, this work addressed the combination of graphite nanoplatelets (GNP) with a soft, tough and crystalline polymer, acting as an efficient binder between nanoplates. With this aim, polycaprolactone (PCL) was selected and exploited in this paper. The crystalline organization of PCL within the nanopaper was studied to investigate the effect of polymer confinement between GNP. Thermomechanical properties were studied by dynamic mechanical analyses at variable temperature and creep measurements at high temperature, demonstrating superior resistance at temperatures well above PCL melting. Finally, the heat conduction properties on the nanopapers were evaluated, resulting in outstanding values above 150 $Wm^{-1}K^{-1}$.

## Experimental part

*Materials*

Polycaprolactone (PCL) is a commercial product purchased from Perstorp UK limited (Capa6500, $M_n$ = 50000, $T_m$ = 56 °C, $T_c$ = 29 °C). Graphite nanoplatelets (GNP) used in this work is supplied by AVANZARE (Navarrete, La Rioja, Spain) prepared via rapid thermal expansion of overoxidized-intercalated graphite, as previously reported[1] and used as supplied without any further treatments. Dimethylformamide (DMF) (99.8%,) purchased from Sigma-Aldrich was used as solvent.

## Preparation methods

Nanopapers were prepared by filtration following the procedure, presented in Figure 1 and described hereunder.

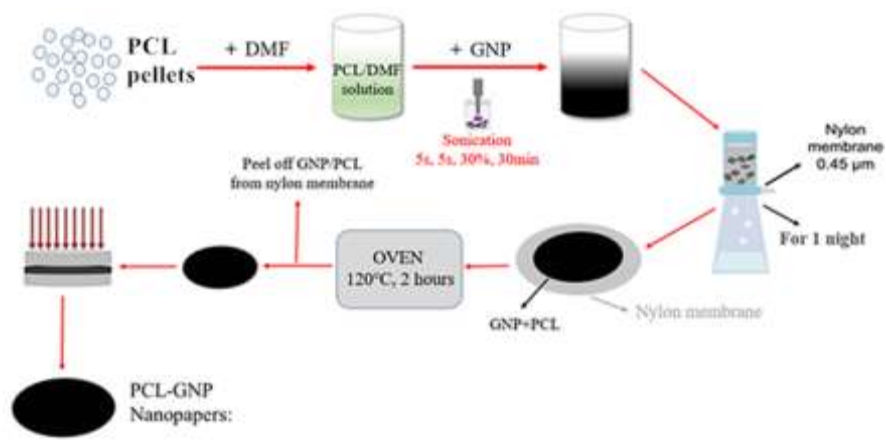

*Figure 1: Preparation procedure of the nanopapers.*

Different amount of PCL pellets (25 mg, 50 mg, 250 mg and 500 mg) were dissolved into 150 ml DMF at 60 °C for 1 hour in order to obtain solutions with different polymer concentrations. GNP powder (50 mg) was added into the prepared PCL solutions. Homogeneous suspensions (no obvious big GNP particles can be seen when transferred to the filter) were obtained by applying a sonication treatment in pulsed mode (5 s on and 5 s off) for 30 min with power set at 30% of the full output power (750 W), accomplished with an ultra-sonication probe (Sonics Vibracell VCX-750, Sonics &Materials Inc.) with a 13 mm diameter Ti-alloy tip. The suspension was transferred into a filtration system equipped with a polyamide supported membrane (0.45μm nominal pore size, diameter 47 mm, Whatman) and left for filtration overnight. After filtration, the cake containing GNP and adsorbed PCL, over the nylon membrane, was dried in two steps, firstly at 70 °C for 2 hours to remove most of the solvent and later at 120 °C for 1 hour to complete solvent removal. Drying in two steps was adopted to avoid cracking of the film, observed when drying in one step at 120 °C, due to the high solvent evaporation rate. Finally, nanopapers were obtained by applying a 6 tons load for 30 minutes on the PCL-GNP cakes after being peeled off from nylon membrane at room temperature (RT). Larger nanopapers were also prepared using 90 mm membrane filters and using 200 mg GNP suspended in 600 ml DMF, while maintaining the same preparation procedure. Hot pressing (80°C and then cooled down to 30°C by water cooling of compression plates) was applied to specimens, to further consolidate the nanopaper structure. Samples codes was defined by indicating the initial ratio of PCL and GNP in the suspensions before filtering, the dimension of the prepared nanopapers and the pressing method, as shown in Table 1.

*Table 1: Nanopapers list, with codes and preparation conditions.*

| Sample code | Ratio PCL : GNP in suspension | Diameter [mm] | Pressing method |
|---|---|---|---|
| PCL10-GNP1-SC | 10 : 1 | 47 | RT |
| PCL10-GNP1-LH | 10 : 1 | 90 | 80°C |
| PCL5-GNP1-SC | 5 : 1 | 47 | RT |
| PCL5-GNP1-LH | 5 : 1 | 90 | 80°C |
| PCL1-GNP1-SC | 1 : 1 | 47 | RT |
| PCL1-GNP1-LH | 1 : 1 | 90 | 80°C |
| PCL1-GNP2-SC | 1 : 2 | 47 | RT |

*Characterization*

Thermal gravimetrical analysis (TGA) was performed with a Mettler-Toledo TGA 1 thermo-gravimetric analyzer. Samples with weight of 5-8 mg were heated from 35 °C to 900 °C under a nitrogen flow of 80 ml/min and then were kept at 900 °C for 20 minutes under oxygen at the same flow rate.

Differential scanning calorimetric (DSC) analysis was performed under a continuous nitrogen purge on a Mettler calorimetric apparatus, model DSC1 STARe/E System. The samples, having a mass between 2.5 and 6 mg, were firstly heated from -10 °C to 200 °C, then cooled down to -100 °C and finally heated to 200 °C again. A scanning rate of 10 °C/min was used on both heating and cooling.

The thermal diffusivity (α) of the prepared nanopapers was measured at 25°C using the xenon light flash analysis (LFA) (Netzsch LFA 467 Hyperflash). The samples were cut in disks with a diameter of 23 mm and the measurements were carried out in a special in-plane sample holder, in which the sample is heated in the central region and the temperature rise was measured on the outer ring of the sample. Measurements were carried out five times for each sample to get an average thermal diffusivity.

Thermal conductivity was calculated from the measured diffusivity values, multiplied by the density and specific heat capacity of the different materials:

$$K = \rho \times \alpha \times C_p \qquad (2)$$

K, thermal conductivity; $\rho$, density of the nanopapers; $C_p$, specific heat capacity of different materials.

The specific heat capacities of nanopapers ($C_{pn}$) were calculated by the weighted average of $C_p$ values of PCL and graphite (0.71 Jg$^{-1}$K$^{-1}$ at RT)[2] for each sample:

$$C_{pn} = C_{pP} \times \Theta_{PCL} + C_{pG} \times (1 - \Theta_{PCL}) \qquad (3)$$

$C_{pP}$, specific heat capacity of PCL, which is 2.0 Jg$^{-1}$K$^{-1}$ at RT;[3] $\Theta_{PCL}$, weight percentage of PCL in the nanopapers; $C_{pG}$, specific heat capacity of graphite.

Thermomechanical properties of nanopapers at different temperatures were investigated by using a Q800 Dynamic Mechanical Analyzer (DMA). The samples were cut into rectangular specimens with dimension of 5×20 mm². The specimen was performed a temperature scan, from room temperature to 150 °C at a heating rate of 2 °C/min, strain of 0.05% and frequency of 1 Hz. Deformation under constant load was carried out at 120 °C under 5 MPa, for 8 hours, followed by deformation recovery at zero load and the same temperature for 8 hours.

## Results and discussion

Composite nanopapers easily obtained by filtration of GNP/PCL suspension demonstrated high flexibility. Indeed, freestanding nanopapers can easily be bent and even folded and then again restored to planar, without breaking, which is impossible for the neat GNP nanopaper, exhibiting remarkable brittleness. As a representative example, pictures for PCL10-GNP1-SC nanopaper are reported in Figure2.

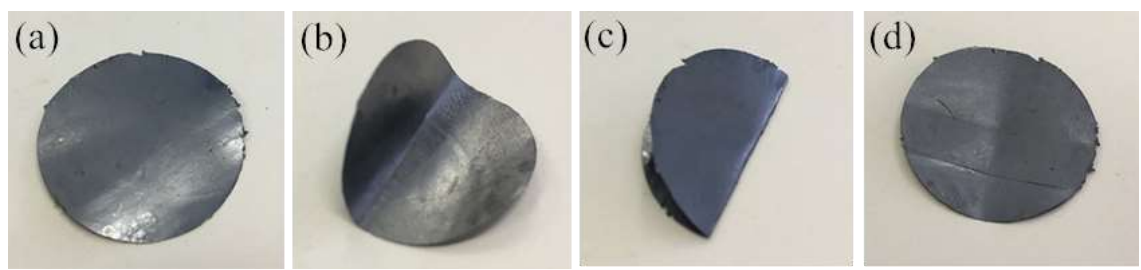

*Figure2: Photographs of freestanding nanopaper PCL10-GNP1-SC: (a) initial nanopaper; (b) nanopaper bent 90º; (c) folded nanopaper; (d) recovery after being bended and folded.*

The morphology of nanopapers in cross-section were investigated by SEM, showing thin deposition of PCL onto the highly oriented GNP flakes. Comparing PCL/GNP nanopapers compressed at room temperature vs the corresponding prepared by hot pressing, significant differences can be found in both thickness and porosity. Indeed, room temperature compressed nanopapers exhibit a higher thickness, typically in the range of 100 μm and delaminated structure. On the other hand, hot pressed counterparts are clearly thinner (approx. 30-40 μm) and more compact, especially for higher PCL/GNP ratio, evidencing the hot pressing stage to consolidate the structure once PCL is above its melting temperature.

The amount of PCL retained by GNP flakes during filtration was investigated by thermogravimetry measurements. Indeed, as PCL has a much lower decomposition temperature ($T_{max}$ at ca. 400 °C) than GNP, it is possible to calculate the polymer content inside the nanopapers from the residual weight at 600°C, as reported in Table 2. The polymer fraction in the nanopapers is clearly much lower than the polymer concentration in the suspension, relative to GNP, demonstrating that only a limited fraction of PCL can be adsorbed onto the GNP flakes and retained in the nanopapers. However, the PCL concentration within the nanopapers is increased by increasing the initial concentration of PCL in the suspensions, relative to GNP. Indeed, ca. 6 wt.% PCL was obtained in PCL1-GNP2-SC whereas concentrations up to about 20 wt.% were obtained for PCL10-GNP1-LH. The PCL content in nanopapers is affected by the initial concentration of the polymers, but it appears to be mainly dependent on the interaction between PCL molecule chains and GNP surface. When the concentration of PCL in the initial suspensions is low, such as PCL1-GNP1 and PCL1-GNP2, the low viscosity of the PCL solution leads to a relatively fast filtration process. When the concentration of PCL solution is gradually increased, the viscosity is increased and this may contribute to retain a higher PCL fraction.

*Table 2: Actual PCL content determined by TGA and enthalpies for endothermic transitions on second heating in DSC. A, B, C and D refers to peaks identified in Figure 3b.*

| Samples | PCL content | ΔH (J/g) of the peaks from second heating stage | | | | |
|---|---|---|---|---|---|---|
| | wt.% | A | B | C | D | Total |
| Neat PCL | 100 | 66.3 | - | - | - | 66.3 |
| PCL10-GNP1-LH | 20 ± 3 | 33.0 | 2.3 | 0.5 | - | 35.8 |
| PCL10-GNP1-SC | 17 ± 3 | 27.2 | 2.8 | 0.9 | 0.8 | 31.7 |
| PCL5-GNP1-SC | 10 ± 1 | 24.0 | 3.6 | 1.0 | 1.2 | 29.8 |
| PCL5-GNP1-LH | 15 ± 3 | 23.4 | 3.4 | 0.8 | - | 27.6 |
| PCL1-GNP1-LH | 7.6 ± 1 | 5.3 | 1.5 | 1.4 | 0.4 | 8.6 |
| PCL1-GNP1-SC | 6.3 ± 0.5 | 4.5 | 2.7 | 1.9 | 4.1 | 13.2 |
| PCL1-GNP2-SC | 6.0 ± 0.6 | 2.0 | 3.0 | 1.8 | 4.8 | 11.6 |

To investigate the organization of PCL chains between GNP, the PCL crystallinity within the nanopapers was addressed, as chain confinement is known to potentially affect crystallinity.[4-5] Beside the fundamental study, crystallinity is also related to the envisaged application of these nanaopapers in heat exchangers. Indeed, crystallinity is one of the most important factors controlling thermal conductivity of polymer materials.[6-9] Crystalline polymers exhibit higher thermal conductivity than amorphous polymers due to the ordered crystal structure, while the random chain conformation in amorphous polymers reduces the phonon mean free path and causes phonon scattering, thus decreasing the heat transfer efficiency.[9-10] The crystallization and melting behaviors of the prepared nanopapers and the neat PCL were characterized by using DSC and results are reported in Figure 3. On cooling plots (Figure 3a) crystallization of pristine PCL can be clearly observed as a sharp peak with max temperature at ca. 28 °C, which is consistent with the well-known crystallization of PCL. On the other hand, the crystallization temperature ($T_c$) for PCL in the presence of GNP raised to ca. 47 °C, i.e. about 20 °C higher than that of the neat PCL, suggesting a significant nucleation activity of GNP flakes on PCL. This crystallization peak is clearly visible for PCL10-GNP1-LH and PCL10-GNP1-SC, while significantly lower and broader signals were obtained for PCL5-GNP1-SC, PCL1-GNP-SC and PCL1-GNP2-SC, which can be partially explained in terms of lower polymer contents (Table 2) within the latter nanopapers.

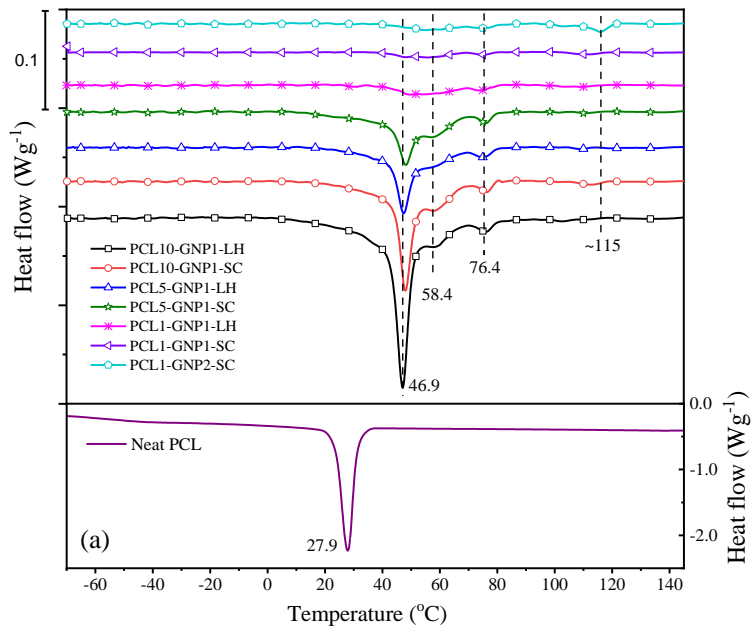

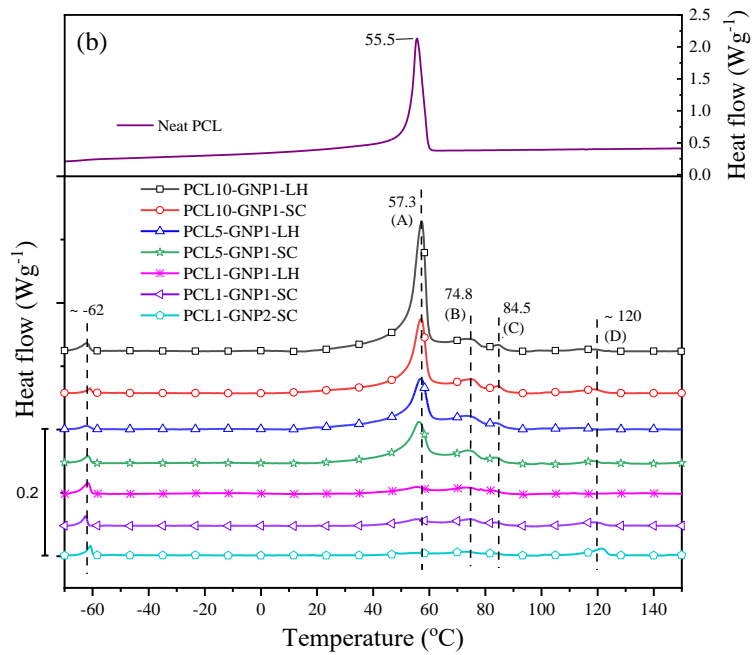

*Figure 3: DSC curves for the cooling (a) and second heating (b) stage.*

The increased $T_c$ for PCL within the nanopapers can be interpreted based on previous literature reports describing strong nucleation activity of graphene-related materials in nanocomposites.[11-14] For PCL, Ahmed et al.[15] reported the effect of GO on the non-isothermal crystallization behavior of PCL, demonstrating an increase in $T_c$ of PCL/GO nanocomposite to ca. 35 °C, compared to ca. 26 °C for the neat PCL, with 1.0 wt.% GO loading. Similar results were reported for PCL/rGO nanocomposite by Wang et al.,[13] with an increase of ca. 10 °C on $T_c$ for the nanocomposite compared to neat PCL. Zhang and coworkers[16] produced nanocomposite based on PCL and thermally reduced graphene oxide (TRGO) and reported $T_c$ of nanocomposite to increase to around 36 °C with TRGO loading of 2 wt.% from 25 °C of neat PCL. Zeng et al.[17] studied the crystallization behavior of PCL/Poly(sodium 4-styrenesulfonate) functionalized GNP (FGNP) composites. Under cooling rate of 10 °C/min, they found that, $T_c$ increase of ca. 8°C and 11°C with addition of 0.05 wt.% and 1 wt.% of FGNP, respectively. A detailed study on non-isothermal crystallization behavior of PCL and PCL nanocomposites with different nanofillers (including GO and graphite powder) and different loadings was done by Kai el at.,[18] the increase in $T_c$ for all the prepared composite being within 10 °C.

Crystallization temperature shift obtained in this work are significantly higher than previously reported for PCL containing graphene related materials, which can be explained by the limited fraction of PCL into the nanopapers, leading to a high interfacial area between GNP and the polymer chains, maximizing nucleation density. Beside $T_c$ shift, it is important to note that extra exothermic peaks at ca. 58, 76 °C and a broad signal above 100 °C were found for all the nanopapers, which did not exist in the case of neat PCL. Relative intensities for these signals, compared to the main crystallization peak, seems to increase when decreasing the total PCL content, thus suggesting such signals to become more important when having little PCL, strongly confined onto GNP flakes.

From the results of second heating, a main endothermic signal in the range between 55 and 58 °C, corresponding to the well-known melting of PCL is clearly observable for both pristine polymer and nanopapers, except for PCL1-GNP2-SC (

Figure 3b). Furthermore, additional signals are found in thermograms for the nanopapers. Indeed a first distinctive features for the nanopapers is found at ca. -62 °C, which is assigned

to the glass transition of PCL.[19] This signal is not visible in pristine PCL, and may therefore suggest a significant fraction of PCL in nanopapers to remain amorphous during the cooling stage. In addition, extra endothermic peaks at ca. 75, 84 and a broad signal around 120 ºC were observed for the composite nanopapers, which were not found for neat PCL, and corresponding to the above described signals for the cooling stage, suggesting the existence of different PCL chains organization. To the best of the authors' knowledge, such high endothermic transitions were never reported for the crystallization of PCL and appear related to a peculiar organization of PCL chains on the surface of GNP. It is worth mentioning that the effect of temperature during nanopapers pressing appears to have some effect on the crystallization behavior of the PCL in nanopapers at relative high temperatures. Indeed, by the comparison of PCL10-GNP1-LH vs. PCL10-GNP1-SC, as well as PCL1-GNP1-LH vs. PCL1-GNP1-SC, smaller enthalpies (Table 2) were found for the PCL high temperature melting peaks, suggesting annealing at 80°C may affect the organization corresponding to the higher transitions.

To quantify the relative amounts of the different crystalline population, the enthalpies of the peaks from second heating stage were calculated, taking into account of the actual PCL contents in nanopapers, and reported in Table 2. The melting enthalpy of the most intense peak (at ca. 57 ºC) was found to increase with increase content of PCL in the nanopapers, which is found to be reverse for peaks at relative high temperatures (peak B, C, D).

Trends for signals A, B and C suggest a strong role of GNP in organizing PCL crystals upon cooling. When a limited amount of PCL present in between GNP, the interaction between PCL chains and GNPs could promote the nucleation process, resulting in higher $T_c$ of PCL. Furthermore, GNPs could also restrict cooperative movements of PCL chains causing a reduction in the total crystallinity of PCL inside the nanopapers. Indeed, the total enthalpy, obtained as the sum of all the peaks for PCL in nanopapers is always lower than in pristine PCL and is found to decrease with decreasing PCL contents.

To investigate the thermomechanical properties of the nanopapers, temperature sweep measurements were performed on hot pressed nanopapers, namely PCL10-GNP1-LH, PCL5-GNP1-LH and PCL1-GNP1-LH by DMTA. These nanopapers demonstrated a significant stiffness at room temperature, with a storage modulus ranging between approx. 7 and 15 GPa, higher stiffness corresponding to lower PCL content, as expected. Interestingly, storage and loss moduli decay vs. temperature is relatively limited and remarkable stiffness are retained for temperature far above the melting of PCL. Indeed, the storage modulus at 150 ºC is about 2.3 and 6.4 GPa of PCL10-GNP1-LH and PCL1-GNP1-LH, respectively, suggesting a very strong adhesion of GNP plates to PCL, even after the polymer melting. The α transition, taken as the maximum of tanδ, is observable at about 90 ºC in all nanopapers, suggesting a remarkable confinement of PCL macromolecules in galleries between GNP flakes.

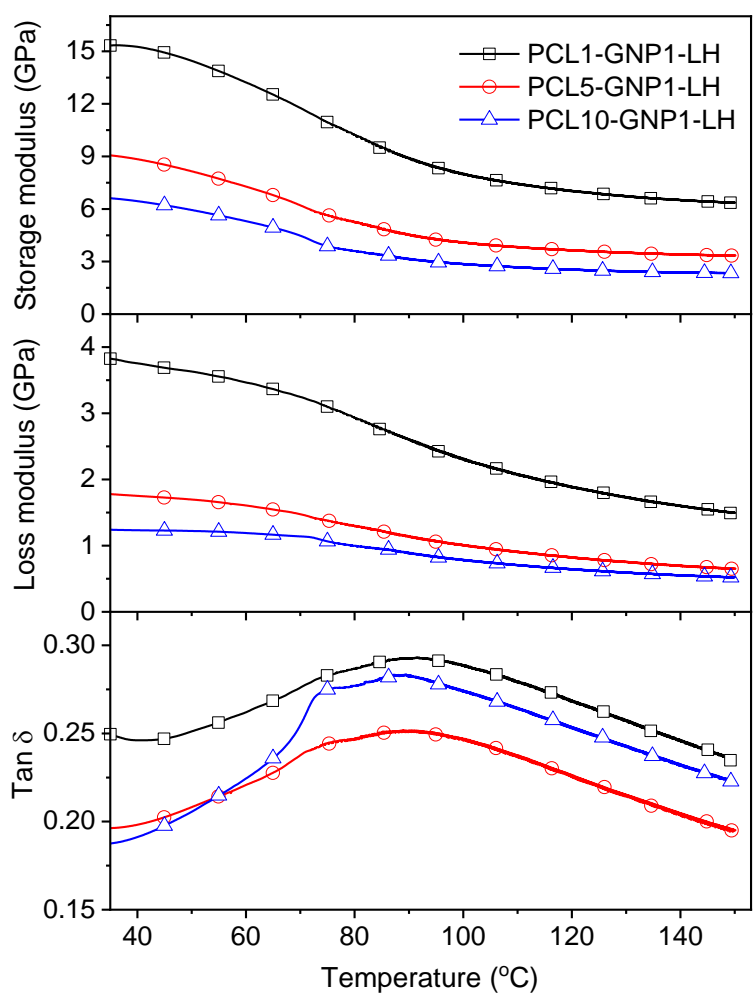

*Figure 4: Temperature sweep DMTA measurement on selected nanopapers.*

The influence of PCL molecule chains on the load-bearing capability of GNPs was further investigated by creep tests. Creep test was carried out at 120 °C under 5 MPa stress, which is representative of operating conditions for low temperature heat exchanger, and result are reported in Figure 5. Upon application of the constant stress, PCL10-GNP1-LH nanopaper immediately deformed to a strain of ca. 2.5% for PCL10-GNP1-LH, followed by a further increase in strain, typical of phase I and II in creep tests, leading to strain of 3.2% after 8 hours creep at 120 °C. After the release of stress, the immediate strain recovery is around 9% of the strain after creep and the final value after 8 hours recovery is close to 12%. Expectedly, creep resistance is even higher for PCL5-GNP1-LH and PCL1-GNP1-LH, owing to the lower PCL content, leading to 2.3 and 0.5% deformation after 8 hours, respectively, which is partially recovered, leading to a final deformation of approx. 1.9 and 0.4%, respectively.

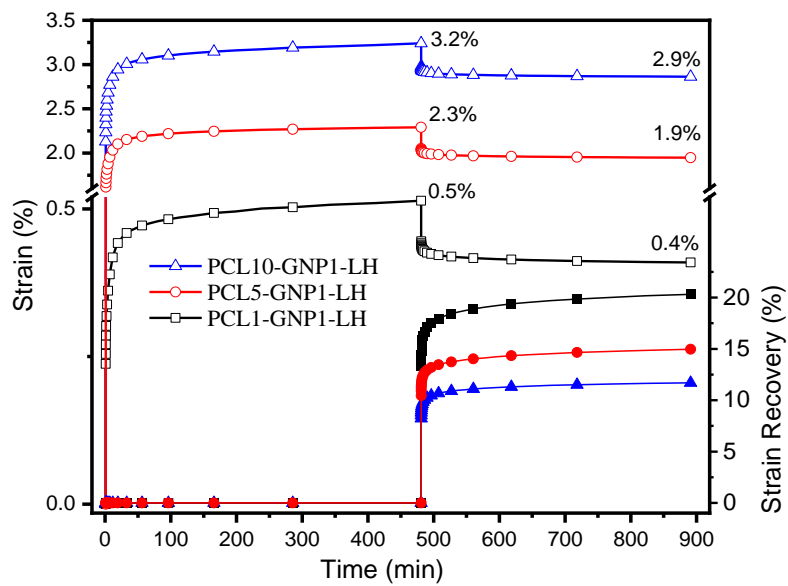

*Figure 5: Strain (empty symbols) and Strain Recovery (filled symbols) plots from creep tests at 120 °C, 5 MPa stress on selected nanopapers.*

These results evidence for outstanding creep resistance of GNP/PCL nanopapers, at temperature far above the melting of PCL, further supporting for the polymer confinement and strong adhesion to GNP.

Envisaging application of these flexible PCL/GNP nanopapers as heat spreaders, thermal diffusivity of the nanopapers was measured. Pristine GNP nanopapers has a thermal diffusivity of 150 ± 3 mm$^2$/s, which may be competitive with traditional metal foils[20-21]. Diffusivity values for the GNP/PCL nanopapers are slightly reduced to 146 ± 2, 127 ± 1 and 138 ± 5 PCL1-GNP1-LH, PCL5-GNP1-LH and PCL10-GNP1-LH, according with the inclusion of a poorly conductive polymer.[22] It is worth noting that reduction in thermal diffusivity are limited and calculation of in-plane thermal conductivity yields outstanding values in the range of 160-190 Wm$^{-1}$K$^{-1}$ (Table 3).

*Table 3: Thermal diffusivity and thermal conductivity values for selected nanopapers*

| Sample | $\Phi_{PCL}$ [wt.%] | Density [gcm$^{-3}$] | Thermal diffusivity [mm$^2$s$^{-1}$] | Thermal Conductivity [Wm$^{-1}$K$^{-1}$] |
|---|---|---|---|---|
| PCL10-GNP1-LH | 20 ± 3 | 1.3 ± 0.1 | 138 ± 5 | 175 ± 16 |
| PCL5-GNP1-LH | 15 ± 3 | 1.4 ± 0.1 | 127 ± 2 | 160 ± 15 |
| PCL1-GNP1-LH | 7.6 ± 1 | 1.4 ± 0.1 | 146 ± 2 | 191 ± 17 |

## Conclusions

The preparation of PCL/GNP nanopapers was carried out to combine thermal and mechanical properties of graphite nanoplated with a soft, tough and crystalline polymer, acting as an efficient binder between nanoplates. Nanopaper characterization evidenced crystallization of PCL is dramatically affected when confined between GNP. Indeed, in addition of the main melting peak, corresponding to pristine PCL, higher temperature transitions were observed, possibly corresponding to higher stability crystals and order-disorder transitions in the organization of PCL chains between GNP. Superior thermal and thermomechanical properties were obtained for PCL/GNP nanopapers, in terms of high viscoelastic modulii, retained up to temperatures well above the melting point of PCL, as well as thermal conductivities above 150 Wm$^{-1}$K$^{-1}$, thus proving prepared materials to bridge the property domains of polymeric materials and conductive ceramics.